\documentstyle[epsf]{europhys}
\def\etal{{\hbox{{\tenit\ et al.\/}\tenrm :\ }}}

\def\And{{\rm and\ }}

\def\stars{\bigskip\centerline{***}\medskip}

\newif\ifboo \boofalse

\def\Review#1{\boofalse{\it #1},}
\def\Name#1{{\sc #1},}
\def\Vol#1{\ifboo Vol. {\bf #1}\else{\bf #1}\fi}
\def\Year#1{\ifboo #1\else(#1)\fi}
\def\Book#1{\bootrue{\it #1},}
\def\Page#1{\ifboo {\rm p. #1}\else{\rm #1}\fi}
\begin{document}
%
%%%   The headers.
%
%%%   These three macros are to have correct headings in your paper.
%%%   You shall omit all the arguments in the two macros `\euro{}{}{}{}'
%%%   `\Date{}' and fill in `\shorttitle{}'. 
%%%   If there is more than one author in the 
%%%   \shorttitle macro, use the macro \etal after first author's name
%%%   to obtain the correct heading.
%
\euro{}{}{}{}
\Date{}
\shorttitle{Y.Y. SUZUKI \etal SIMULATION OF POLYMERS IN A CURVED BOX}
%
%%%  The title, the Author(s) and the affiliation(s)
%
%%%   The title is set in bold (initial word only is capitalized).
%%%   Mathematical expressions and formulas within the title shall be left
%%%   in light face. Initial(s) of the first name(s) are followed by the
%%%   author(s)'s last name(s). If the authors have different affiliations,
%%%   the name must be followed by one or more \inst{number} each referring
%%%   to one of the addresses to appear in the following macro \institute.
%%%   Other items like `Present address' or `email' may be added by putting
%%%   a `\footnote' after the last \inst{number}.
%%%   Begin each address with \inst{number}; the end of an address is \\;
%%%   \\ can also be used to break a line.
%
\title{Simulation of polymers in a curved box:
Variable range bonding models}
\author{Y.Y. Suzuki\inst{1}\footnote{E-mail address: suzuki@brl.ntt.co.jp},
      T. Dotera\inst{2}
	\And M. Hirabayashi\inst{2}}
\institute{
     \inst{1} NTT Basic Research Laboratories, Atsugi 243-0198, Japan\\
     \inst{2} Saitama Study Center, the University of the Air, 682-2 
	Nishiki-cho, Omiya 331-0851, Japan}
%
%%%    The `\maketitle' macro needs the following macro:    \rec{}{}
%%%    to be left empty.
%
\rec{}{}
%
%%%   Physics Abstracts Classification.
%
%%%   There are two macros: the first one `\pacs{}' makes the PACS 
%%%   environment,the second one `\Pacs{}{}{}' can be used for each
%%%   classification you need.
%%%   To create the subject index of the volume it is important to divide
%%%   the classification numbers into the three different arguments like
%%%   in the following examples 
%
\pacs{
\Pacs{36}{20$-$r}{Macromolecules and polymer molecules}
\Pacs{05}{10Ln}{Monte Carlo methods}
\Pacs{82}{65Pa}{Surface-enhanced molecular states and other
gas-surface interactions}
      }
\maketitle
%
%%%   ! Don't forget this command to format the title page of your article!
%
%%%   The Abstract
%
\begin{abstract}
We propose new polymer models for Monte Carlo simulation and apply
them to a polymer chain confined in a relatively thin box which has
both curved and flat sides, and show that either an ideal or an
excluded-volume chain spends more time in the curved region than in
the flat region.  The ratio of the probability of finding a chain in
the curved region and in flat region increases exponentially with
increasing chain length.  The results for ideal chains are
quantitatively consistent with a previously published theory.  We find
that the same effect appears with excluded-volume chains and a similar
scaling relation can be applied to them up to a certain length of the
polymer.
\end{abstract}
%
%
%%%   Main text
%
%%%   Sectioning
%
%%%   In EuroPhys there is only ``one'' level of sectioning `\section{}'.
%
\section{Introduction}
\begin{figure}
  \epsfysize=4cm
  \centerline{\epsfbox{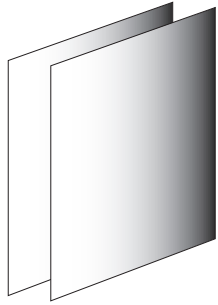}}
\caption{flat box}
\label{flat}
\end{figure}

The structure~ \cite{casassa}, \cite{daoud}, \cite{brochard},
\cite{wyart} and dynamics~\cite{kremer}, \cite{milchev}, \cite{hagita}
of polymer chains confined within a narrow space such as in a slit
(Fig.~\ref{flat}) are important for practical applications, e.g.
filtration, gel permeation chromatography and oil recovery, and have
attracted much interest.~\cite{deGennes} The effect of
confinement on polymer chains was calculated theoretically in
well-defined geometries (a slit and a tube) and was simulated
by a lattice model~\cite{wall} and an off-lattice model.~\cite{kumar}

Recently, polymer confinement in curved boxes (a cylindrical shell, an
undulated slit, etc) was studied.~\cite{yaman}, \cite{dotera} Such
studies can relate to not only the behaviour of polymers confined in
complex rigid geometries but also the behaviour of deformable bilayer
membranes confining the polymer chains.

\begin{figure}
  \epsfysize=4cm
  \centerline{\epsfbox{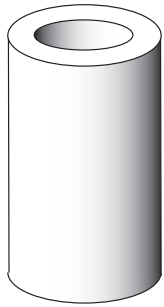}}
\caption{cylindrical shell}
\label{cylinder}
\end{figure}

Yaman {\it et al.}~\cite{yaman} showed that an ideal (Gaussian)
polymer chain confined between cylindrical shells
(Fig.~\ref{cylinder}) has a lower free energy than one confined
between two flat surfaces (Fig.~\ref{flat}), and predicted that
polymer chains confined in bilayer membranes reduce the effective
bending rigidity of the membranes and might induce spontaneous
curvature in the system, e.g. leading to transitions from lamellar to
bicontinuous phases.

Above mentioned studies of the confinement effects are limited to
ideal chains.  The ideal chain represents the polymers in $\Theta$
solvents where the excluded-volume interactions are compensated for by
monomer-monomer interactions.  It can act as an unperturbed model for
a perturbation calculation of the ``real'' chain in a good solvent
characterized by excluded-volume interactions.  However, the
perturbation calculation is difficult for confined flexible polymers
and few publications have addressed the excluded-volume
effects.~\cite{dietrich}

In this paper, we investigate both curved surface effects and
excluded-volume effects on a flexible polymer confined between
two surfaces using Monte Carlo simulations and compare with an ideal
chain confined in the same geometry.  We suppose that the surfaces are
purely repulsive (no trend towards adsorption), with and without
excluded-volume interactions.  In particular, we are interested in the
polymer behaviour when the polymer chain has a large size relative
to the gap between the confining surfaces.  The effect is sensitive to
the curvature of the surfaces, therefore, our off-lattice simulation
has the strong advantage over a lattice simulation where the curved
surface is described by a zig-zag boundary (steps of flat surface
strips).

\begin{figure}
  \epsfysize=4cm
  \centerline{\epsfbox{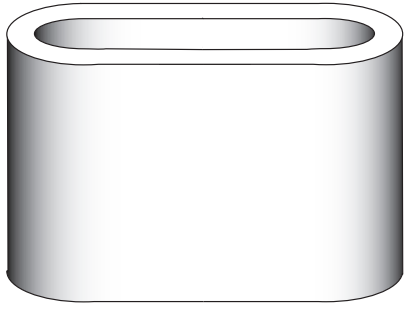}}
\caption{race-track box}
\label{track}
\end{figure}

First, we review the analytical study for ideal chains by Yaman {\it
et al.}~\cite{yaman}  Next, we propose new simulation models and check our
algorithms in the free $n$-dimensional spaces to see if the models
reproduce the known characteristics of flexible polymers.  Then, we apply
our simulation to ideal chains confined in a race-track box composed
of curved and flat regions (Fig.~\ref{track}) in order to compare with
the analytical results by Yaman {\it et al.}  Finally, we
apply our simulation to excluded-volume chains confined to the
race-track box.

\section{Analytical study for ideal chains by Yaman et al}
For a random walk of $N$ steps with endpoints at $r$ and $r'$, the
propagator $G(r,r',N)$ is defined as \cite{deGennes}
\begin{equation}
\left( \frac{\partial}{\partial N} - \frac{l^2}{6} \nabla_r^2  \right) G(r,r',N)= \delta(r-r') \delta(N),
\end{equation}
where $l$ is the length of one step.  The ideal chain is modelled as a
random walk in space, where $l$ and $N$ correspond to the
characteristic length of a monomer and the degree of polymerizations,
respectively.

The partition function is a sum over all configurations:
\begin{equation}
Z=\frac{1}{V}\int dr dr' G(r,r',N).
\end{equation}
With the eigenfunction expansion: 
\begin{equation}
G(r,r',N)=\sum_n \Psi_n^*(r) \Psi_n(r') e^{-N E_n l^2/6} \Theta (N),
\end{equation}
where $\Theta (N)$ is the step function, our problem of a polymer
confined in a spherical annulus in arbitrary space dimension $D$ with
radius $a$ and thickness $d$ reduces to
\begin{equation}
\nabla^2 \Psi(r)=-E\Psi(r),
\end{equation}
under SO(D) symmetry and the boundary conditions:
\begin{equation}
\Psi(|r|=a)=\Psi(|r|=a+d)=0.
\end{equation}

For the radial wave function:
\begin{equation}
\left\{ \frac{d^2}{du^2}+(D-1)\frac{1}{u} \frac{d}{du}+\epsilon-\frac{\xi^2}{u^2} \right\} \Phi_{\nu}(u)=0, \label{eq:wave}
\end{equation}
where $\epsilon=Ed^2$, $\xi^2=\nu^2+(D-2)^2/2$, and $u=r/d$, the
boundary conditions are $\Phi_{\nu}(\eta)=\Phi_{\nu}(\eta+1)=0$, where
$\eta=a/d$.  The general solution of Eq.(\ref{eq:wave}) is
\begin{equation}
\Phi_{\nu}(u)=A J_{\xi}(\sqrt{\epsilon} u) + B N_{\xi}(\sqrt{\epsilon} u),
\end{equation}
where $J$ and $N$ are the standard Bessel functions.  The boundary
condition determines the relation between $\epsilon$ and $\eta$.  For
spherical symmetry, only $\nu=0$ states contribute to the
partition function.  We label these $s$-wave states by $n=1,2,\dots$,
and define the normalized radial wave function as $R_n(r)$ for
$\nu=0$.

For a spherical annulus with a large inner radius $a$ and a small
thickness $d$, $E_n$ can be obtained perturbationally:
\begin{equation}
E_n(a,d) = \left( \frac{n \pi}{d} \right)^2 + \frac{(D-1)(D-3)}{4d^2} 
\left[ \left( \frac{d}{a} \right)^2 -\left( \frac{d}{a} \right)^3 + \left(  \frac{d}{a} \right)^4 \left( 1- \frac{3}{8n^2 \pi^2} \right) + \cdots \right].
\end{equation}
Note that $D=1$ is a slit between two flat surfaces (Fig.~\ref{flat}),
$D=2$ is a cylindrical shell (Fig.~\ref{cylinder}), $D=3$ is a
spherical shell.  The curvature lowers the
spherically symmetric energy spectrum for $1<D<3$.

The partition function becomes
\begin{equation}
Z=\frac{1}{V}\sum_{n=1}^{\infty} e^{-N E_n(a,d) l^2/6} \left| \int dr r^{(D-1)}R_n(r) \right|^2.
\end{equation}
To indicate explicitly the dependence of the energies on $a$ and $d$, we introduce $B_n(a,d)$:
\begin{equation}
B_n(a,d) \equiv \frac{\left| \int dr r^{(D-1)}R_n(r)  \right|^2}
{\frac{S_{D-1}}{D} \left\{ (a+d)^D -a^D \right\} }
\end{equation}
where $S_D$ is the surface area of the unit sphere in $D$-dimensional space.
The free energy is then given by
\begin{equation}
F= -k T \ln \left[ \sum_{n=1}^{\infty} B_n(a,d) e^{-E_n(a,d)Nl^2/6} \right].
\end{equation}

In this paper, we are interested in the case of a cylindrical shell
($D=2$) in three dimensions, for a polymer size $R_g \sim \sqrt{N} l$
larger than the shell thickness $d$.  Then, the leading term of the
free energy difference between the cylindrical shell and the flat box
($D=1$) is
\begin{equation}
\Delta F \sim kT \left( E_1 |_{D=2} - E_1 |_{D=1} \right) \frac{N l^2}{6} = -kT N\frac{l^2}{24a^2}.
\label{eq:deltaF}
\end{equation}

From this result, Yaman {\it et al}.\ predicted that confining an
ideal polymer chain into the bilayer membrane lowers the effective
bending rigidity of the membrane and might induce spontaneous symmetry
breaking of the membrane.  Another interpretation of this result is
that a polymer chain confined in the rigid shell is attracted into
curved region, and that the ratio of probability finding a polymer chain
in equivalent volume of curved and flat regions should be
\begin{equation}
C=e^{- \Delta F/kT}=\exp \left( \frac{Nl^2}{24a^2} \right).
\label{eq:C}
\end{equation}
Here, we assume that the size of the confined polymer is smaller than
the characteristic length of the curved and flat sections.

\section{Models and method}
Our polymer simulations are off-lattice beads and bonds models with $N+1$
beads and $N$ bonds; they are relaxed versions of the random flight
model with variable bond lengths.  We call this the variable range bonding
model.  The models consist of the following conditions: (1)
Connectivity: $|\mbox{\boldmath{$r_i-r_{i+1}$}}| \leq 1.0$,
where $ \mbox{\boldmath$r_i$} = (r_{i1},r_{i2},\cdots,r_{in})$ stands
for the position of the $i$-th bead.  (2) Excluded-volume: $
|\mbox{\boldmath{$r_i-r_j$}}| \geq 2 r_e, $ for all $i,j$.  $r_e$ is
the radius of the excluded volume of the beads.  (3) Geometric constraint: the beads are in a confining geometry, here a race-track
box (Fig.~\ref{track}).  Note that we impose the last condition only
on beads, but not on bonds.  Thus, we even allow a bond that cuts
through the curved wall as long as two beads attached to the bond
satisfy the geometric constraint.  We consider two models: Model-I
which requires (1) and (3), and Model-E which requires (1)--(3).  They
correspond to the ideal polymer and the excluded-volume polymer
respectively.

Our Monte Carlo (MC) algorithm is the following.  i) Select a bead at
random.  ii) Generate a trial jump: $ \mbox{\boldmath{$r$}}
\rightarrow \mbox{\boldmath{$r + \Delta$}} $.  Each component of the
jump vector $\Delta_{\mu}$ ($\mu=1,2,\cdots,n$) is independently
determined by the Gaussian distribution with zero mean value and a
standard deviation of 0.3 for Model-I and 0.15 for Model-E.
The Gaussian distribution is generated by the
Box-Muller method.~\cite{press} iii) Check if the new position satisfies
above mentioned conditions: (1) and (3) for Model-I, (1)--(3) for
Model-E; if it does, accept the jump; if not, abandon the jump.  iv)
Return to i).  For a simulation with $N+1$ beads, $N+1$ cycles
comprise one MC step.

The reason to halve the standard deviation of the Gaussian
distribution of the jump distance for Model-E is to increase
the acceptance ratio of the bead jump although the bead in
Model-E is restricted by the excluded-volume interactions compared
with Model-I.

For Model-I in free $n$-dimensional space, the bond distribution
is expected to be uniform inside the $n$-dimensional sphere with
radius 1.  Therefore, the bond length distribution is described by the
distribution function $P(r)$ of bond length $r$ for $0 \leq r \leq 1$
as
\begin{equation}
P(r) = n r^{n-1}, 
\end{equation}
and
\begin{equation}
\int_0^1 P(r) dr = 1.
\end{equation}
Thus, the mean bond lengths are analytically calculated as 
\begin{equation}
\int_0^1 r^2 P(r) dr = n/(n+2),
\end{equation}
and the chain should be a Gaussian chain with bond length
$\sqrt{n/(n+2)}$.  The mean square end-to-end distance is analytically calculated as $\langle R_N^2 \rangle = N n/(n+2)$.  For 3-dimensions, the mean
square bond length becomes $\langle l^2 \rangle=0.6$.  To evaluate
the mean square end-to-end distances $\langle R_N^2 \rangle$
by simulations, we made ten runs of simulation with $4 \times 10^6$ MC steps for each
length up to 200 beads.  The simulation results are linearly fit by:
$\langle R_N^2 \rangle = 0.334 N$ in 1D, $0.492 N$ in 2D, $0.594 N$ in
3D, which implies that $\langle l^2 \rangle=0.594$ for 3-dimensions.
They are consistent with the analytical calculations.

\begin{figure}
  \epsfysize=4cm
  \centerline{\epsfbox{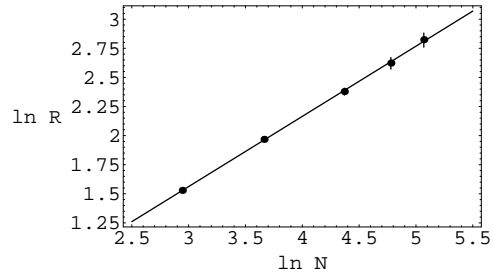}}
\caption{Log-log plot of the end-to-end distance of Model-E in free 3D
space.  The fitting curve is $R \propto N^{0.601}$.}
\label{free}
\end{figure}

For Model-E in the free $n$-dimensional space, we also carried out a
series of simulations with $4 \times 10^6$ MC steps.  We plot the
results for free 3-dimensional space in Fig.~\ref{free}.
The fitting exponent $\nu=0.601$ matches the best theoretical
evaluation $\nu=0.588$ with 2 \% error.  The square average of the
bond length is obtained as $\langle l^2 \rangle=0.6742$.

Our race-track geometry (Fig.~\ref{track}) consists of two flat
sections and two curved sections.  The width of the track is $d=0.5$,
the radius of the inside wall of the curved sections is $a=4.75$, and
the length of the flat sections is $L=15.71$.  The volume of the
curved region and the flat region are the same.

For race-track simulations of Model-E, we use the same geometry as
used for the Model-I, except that the width of the race-track $d=0.5$
is defined relative to the center of the beads.  When the width is
defined relative to the surface of the beads, it is $d+2r_e=1.0$ and the radius of the inside wall is $a-r_e=4.5$
where $r_e=0.25$ is the radius of the beads.

To avoid localization on one side of the track, the polymer is relocated,
every $10^6$ MC steps, to the position of $180^\circ$ rotation with
respect to the center point of the track.
The position data of beads are gathered every 1000 MC
steps.  Such data give us an idea of a density profile of the polymer
in the restricted geometry.  We divide the number of beads recorded in
the curved region by the number of beads recorded in the flat region
to obtain the probability ratio $C$.

The advantages of our polymer models are 1) the movement of polymer in
the narrow space is facilitated due to variable bond lengths and
variable range bead jumping, and 2) the high probability of polymer
approaching to the wall compared to models with fixed bond length and
fixed range bead jumping is convenient to observe the effects in
proximity to the curved wall.

\section{Results and discussions}
For each model and each polymer length, we carried out 5 runs of simulation 
with $4 \times 10^8$ MC steps after an initial randomized stage ($10^6$
MC steps).

\begin{figure}
  \epsfysize=4cm
  \centerline{\epsfbox{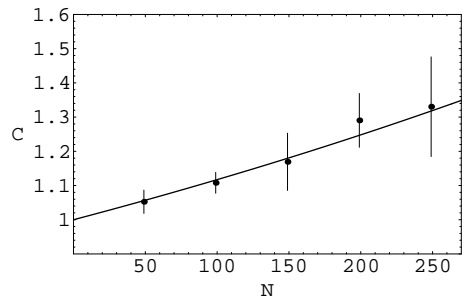}}
\caption{Results of Model-I.  The line is drawn according to the
theoretical curve with parameters $a=4.75$, $\langle l^2 \rangle=0.6$: $C=\exp
(N/902.5)$.}
\label{gaussian}
\end{figure}

For Model-I, the race-track simulation is performed for each chain
length (number of beads) : $N+1=$ 50, 100, 150, 200, 250.  The results
are plotted by in Fig.~\ref{gaussian}.  The geometry gives parameters:
$d=0.5$ and $a=4.75$.  The mean square of the bond length is analytically
calculated as $\langle l^2 \rangle=0.6$.  Then, the theoretical curve
with the parameters is given by $C=\exp (N/902.5)$.

\begin{figure}
%
%\vbox to 1cm{\vfill\centerline{\fbox{Here is the figure 3}}\vfill}
%
  \epsfysize=4cm
  \centerline{\epsfbox{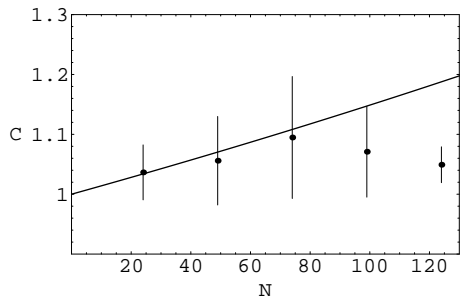}}
\caption{Results of Model-E.  The line is drawn according to the
theoretical curve for a Gaussian chain with parameters
$a=4.5$, $\langle l^2 \rangle=0.6742$: $C=\exp (N/720.9)$.}
\label{excluded}
\end{figure}

For Model-E, the race-track simulation is performed for each chain
length: $N+1=$ 25, 50, 75, 100, 125.  The results are plotted in
Fig.~\ref{excluded}.  We try to fit the data of the excluded-volume
chains by the theoretical curve developed by Yaman {\it et al}.\ for
the Gaussian chain.  The geometry gives parameters: $d=1.0$ and $a=4.5$.
The mean square of the bond length is evaluated by our free space
simulations as $\langle l^2 \rangle=0.6742$.  Then, the theoretical curve with the parameters is given by $C=\exp (N/947.5)$.

In Model-I, our simulation results (Fig.~\ref{gaussian}) fit the
analytical formula developed by Yaman {\it et al}.\ for all the chain
lengths.  In Model-E, our results (Fig.~\ref{excluded}) are well
described by the same formula as far as the chain lengths are short,
however, they show a deviation when the chain gets longer.

The excluded-volume effects result in a elongated equilibrium polymer
shape compared with the ideal chain.  This is confirmed by the snap
shots of our simulations.  They show that the length of the
excluded-volume chain with $N+1=75$ confined in the track is
comparable to the section length $L \sim \pi a$.  If the chain is too
long to be accommodated in a single section, our interpretation of the
free energy as a probability is not justified.  Beyond this length, a
single curved section cannot accommodate the whole chain and the chain
cannot squeeze into the section due to the excluded-volume effects.
Therefore, a deviation starting beyond this length is consistent with
our analysis. The probability ratio $C$ should reach a maximum in the
vicinity of this length.  This is consistent with our data.

\section{Conclusions}

We present new polymer models for Monte Carlo simulation which work
well for our purpose.  In particular, they have the advantage of efficient
changes of the polymer conformation in restricted geometries due
to the variable range character of bond lengths and of jump distances.

We have investigated the curvature effect on a flexible polymer
confined in a thin box.  We have plotted the ratio of probability $C$
finding a polymer between in a curved region and in a flat region as a
function of chain length $N$.  For the ideal chains, our results agree
quantitatively with the prediction of the analytic theory.  We also
observe a similar curvature effect on excluded-volume chains whose
lengths shorter than the length of the curved section of the box.

Our simulations only cover the linear region of $C$ as a function of
chain length $N$, due to the limitation of our computer power for the
case of model-I and due to restriction of geometrical scale of
confinement for the case of model-E.  However, our results may still
be useful under present circumstances where there is no
experimental data for either ideal or excluded-volume chains and there is no
analytical theory for excluded-volume chains.

%For the further development of our study, the authors are working on
%an intuitive explanation of the curvature effects and an analytical
%formula for the curvature effect on excluded-volume chains.

%
%%%   In the acknowledgments, use the following macro  before and instead 
%%%   of  ``Acknowledgments''
%
\stars

The authors thank F. Wittel, P. Pincus and K. Kremer for useful
suggestions.  Part of this work was done during the stay of one of the
authors (Y.Y.S.) at MRL UCSB.

%%%   Bibliography environment begins here. You can use the macros \Name{},
%%%   \And, \Book{} or \Review{}, \Vol{}, \Year{} and \Page{}, to type your
%%%   references.
%

\end{document}